\documentclass{PoS}
\usepackage{comment}
\usepackage{graphicx}
\usepackage{enumitem}
\usepackage[sorting=none,style=numeric]{biblatex}
\bibliography{mybib} 

\title{An Indirect Dark Matter Search Using Cosmic-Ray Antiparticles with GAPS}

\ShortTitle{The General Antiparticle Spectrometer}
\author{\speaker{Alexander Lowell} for the GAPS Collaboration\\
        University of California, San Diego\\
        E-mail: \email{awlowell@ucsd.edu}}

\abstract{Experiments aiming to directly detect dark matter (DM) particles have yet to make robust detections, thus underscoring the need for complementary approaches such as searches for new particles at colliders, and indirect searches of DM decay or annihilation signatures in photon and cosmic-ray spectra. In particular, low energy (< 0.25 GeV/n) cosmic-ray antiparticles such as antideuterons are strong candidates for probing various DM models, as the yield of these particles from DM processes can exceed the conventional astrophysical background by up to two orders of magnitude. The General Antiparticle Spectrometer (GAPS), a balloon borne cosmic-ray detector, will exploit this idea and perform a virtually background-free measurement of the cosmic antideuteron flux in the regime < 0.25 GeV/n, which will constrain a wide range of viable DM models. Additionally, GAPS will detect approximately 1000 antiprotons in an unexplored energy range throughout one long duration balloon (LDB) flight, which will constrain < 10 GeV DM models as well as validate the GAPS detection technique. Unlike magnetic spectrometers, GAPS relies on the formation of an exotic atom within the tracker in order to reliably identify antiparticles. The GAPS tracker consists of ten layers of lithium-drifted silicon detectors which record dE/dx deposits from primary and nuclear annihilation product tracks, as well as measure the energy of the exotic atom deexcitation X-rays. A two-layer, plastic scintillator time of flight (TOF) system surrounds the tracker and measures the particle velocity, dE/dx deposits, and provides a fast trigger to the tracker. The nuclear annihilation product multiplicity, deexcitation X-ray energies, TOF, and stopping depth are all used together to discern between antiparticle species.  This presentation provided an overview of the GAPS experiment, an update on the construction progress of the tracker and TOF systems, and a summary of the expected performance of GAPS in light of the upcoming LDB flight from McMurdo Station, Antarctica in 2020.}

\FullConference{The 39th International Conference on High Energy Physics (ICHEP2018)\\
		4-11 July, 2018\\
		Seoul, Korea}

\begin{document}

\section{Introduction}
\begin{figure}[t]
\begin{minipage}[t]{0.49\linewidth}
\includegraphics[width=\linewidth]{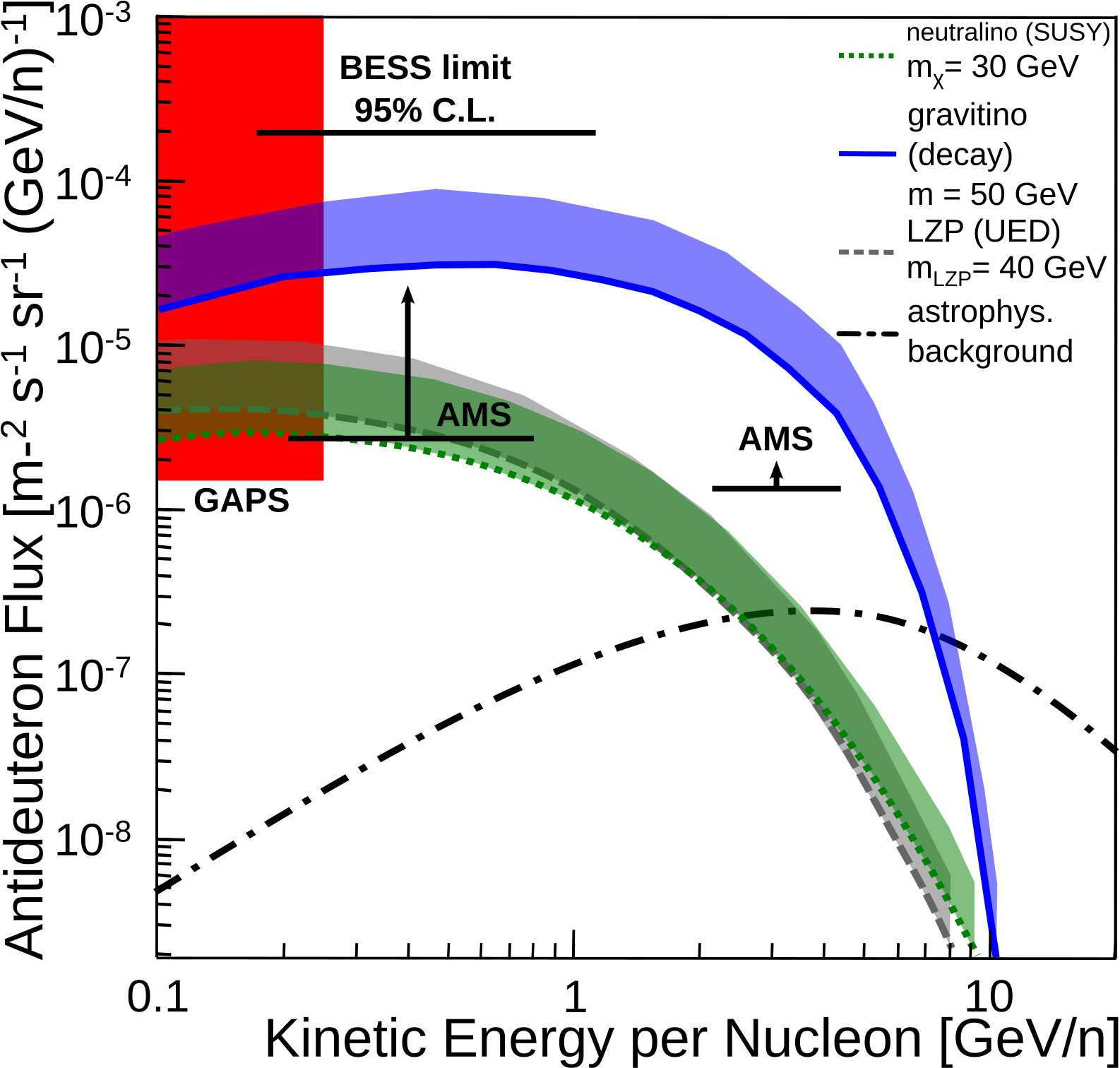}
\caption{Antideuteron fluxes from three representative WIMP models, as well as the astrophysical background component.  The bands reflect the uncertainty in propagation modeling.  Simulated sensitivities are shown for GAPS (100 days of flight) and the Alpha Magnetic Spectrometer (AMS, 5 years in orbit), with arrows indicating the size of the geomagnetic cutoff correction for the AMS orbit.  The 95\% confidence upper limit from BESS is also shown \cite{fuke2005}.  GAPS will improve the BESS limit by $\sim2$ orders of magnitude, and achieve a comparable sensitivity to AMS in a fraction of the observation time.}
\label{fig:dbar}
\end{minipage}
\hfill
\begin{minipage}[t]{0.49\linewidth}
\includegraphics[width=\linewidth,trim={0 -2cm 0 0}]{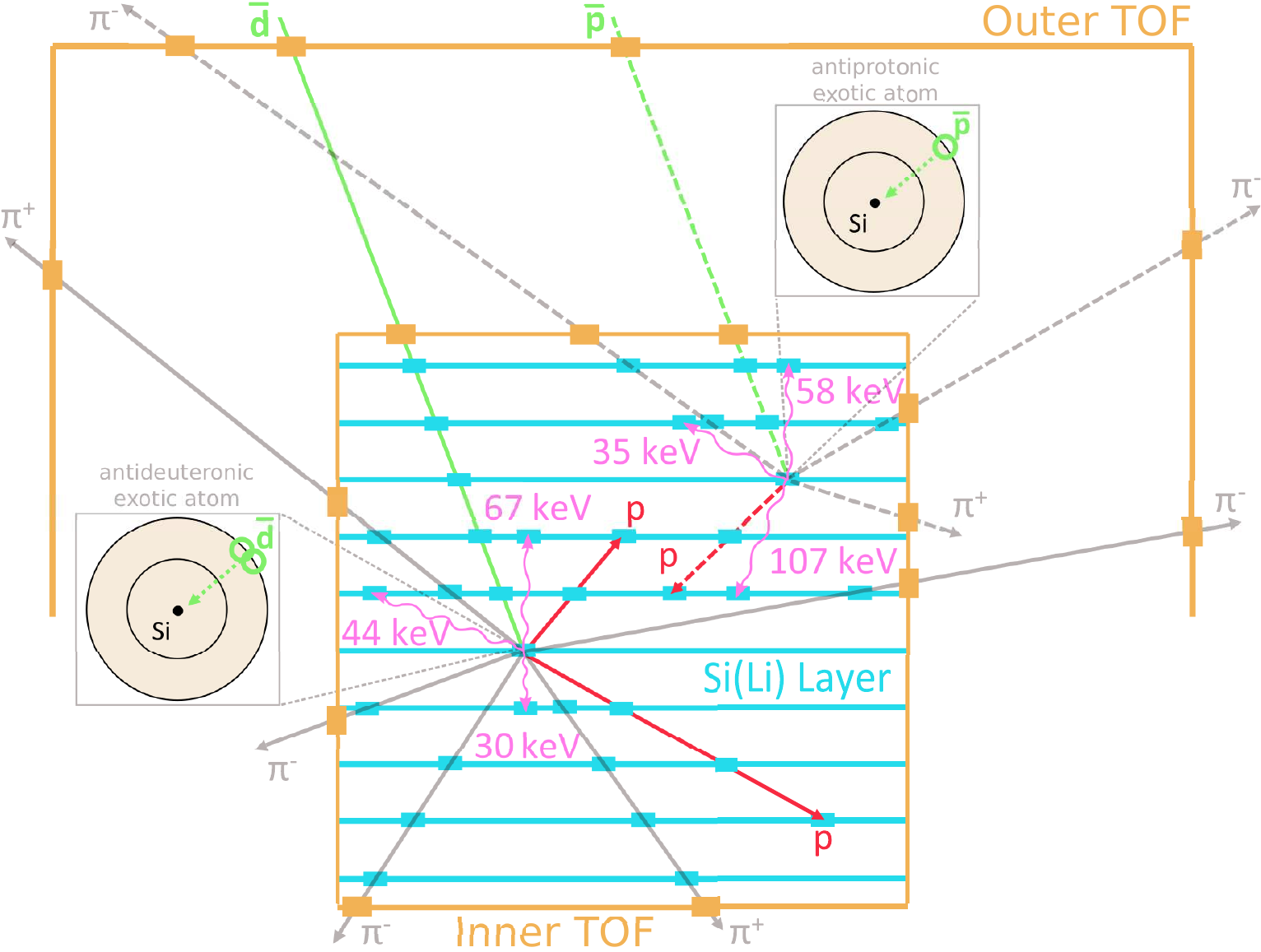}
\caption{An incident cosmic-ray antinucleus (green) strikes the outer TOF, inner TOF, and is subsequently stopped in the silicon tracker system.  Before annihilating, the antinucleus enters a bound state with a Si nucleus in the tracker (insets), thus forming an exotic atom in a highly excited state.  The exotic atom de-excites, emitting Auger electrons and characteristic X-rays (purple) which are absorbed in nearby detectors.  The annihilation finally takes place resulting in a ``star'' of pions (grey) and protons (red), which further deposit energy in both the tracker and TOF detectors.}
\label{fig:gapscartoon}
\end{minipage}%
\end{figure}

Astronomical observations have confirmed the existence of dark matter in our universe through its gravitational influence on visible matter, yet its particle nature remains a mystery.  A well-motivated explanation of the DM phenomenon is that the DM is made up of weakly interacting massive particles (WIMPs), currently existing outside the reach of the standard model (SM).  In a variety of WIMP scenarios, WIMPs may self-annihilate or decay and produce SM particles and antiparticles in equal amounts.  Detecting these SM particles indirectly probes DM processes in the Milky Way, and several results in recent years have underscored the value of such indirect searches (see \cite{conrad2017} for a review).  Additionally, indirect searches can be sensitive to certain types of WIMPs - such as the decaying gravitino \cite{dal2014} - that would be inaccessible to direct detection experiments.  Low energy (< .25 GeV/n) antideuterons in cosmic rays present a unique opportunity for indirect DM searches, due to the kinematic suppression of low-energy antideuteron production in cosmic-ray collisions (see \cite{aramaki2016} for a review).  Figure \ref{fig:dbar} illustrates the unique advantage of low-energy antideuterons, where the signal to background flux ratio can easily exceed $\sim 100$ in the range $0.1-0.25$ GeV/n for several well-motivated WIMP DM models.  Despite this, antideuterons have yet to be observed in cosmic rays due to their low absolute flux, and the high required rejection power against more abundant species with similar event characteristics such as antiprotons.

\section{The GAPS Instrument}

The General Antiparticle Spectrometer (GAPS) is a balloon-borne detector of cosmic-ray antinuclei, optimized for the detection and identification of low-energy antideuterons and antiprotons using an exotic atom technique.  A cartoon illustrating the GAPS operating principle is shown in Figure \ref{fig:gapscartoon}.  Annihilating antinuclei in the GAPS detectors produce a rich event topology with many features for particle identification and background rejection, including but not limited to:
\begin{itemize}[noitemsep,topsep=0pt]
\item Stopping depth and dE/dx energy loss.  An antideuteron will penetrate approximately twice as deep into the tracker as an antiproton with the same time of flight.  For equal stopping depths, an antideuteron will deposit more energy per layer than an antiproton.
\item Characteristic X-ray energies.  The atomic shell structure of an antideuteronic exotic atom is different than that of an antiprotonic atom.  Therefore, the energy of the deexcitation X-rays can be used to identify the exotic atom type.
\item Pion and proton multiplicity.  The number of pions and protons generated in the annihilation scales with the number of antinucleons, and can therefore be used as a feature for antinucleus identification.
\end{itemize}
While antiprotons are the dominant background for the antideuteron measurement, they are also messengers of light DM (< 10 GeV) processes \cite{aramaki2014}, and will serve to validate the exotic atom technique in flight.  GAPS will detect approximately 1000 antiprotons during 30 days of flight.

\subsection{Lithium Drifted Silicon Tracker}

\begin{figure}[t]
    \centering
    \includegraphics[width=0.75\textwidth]{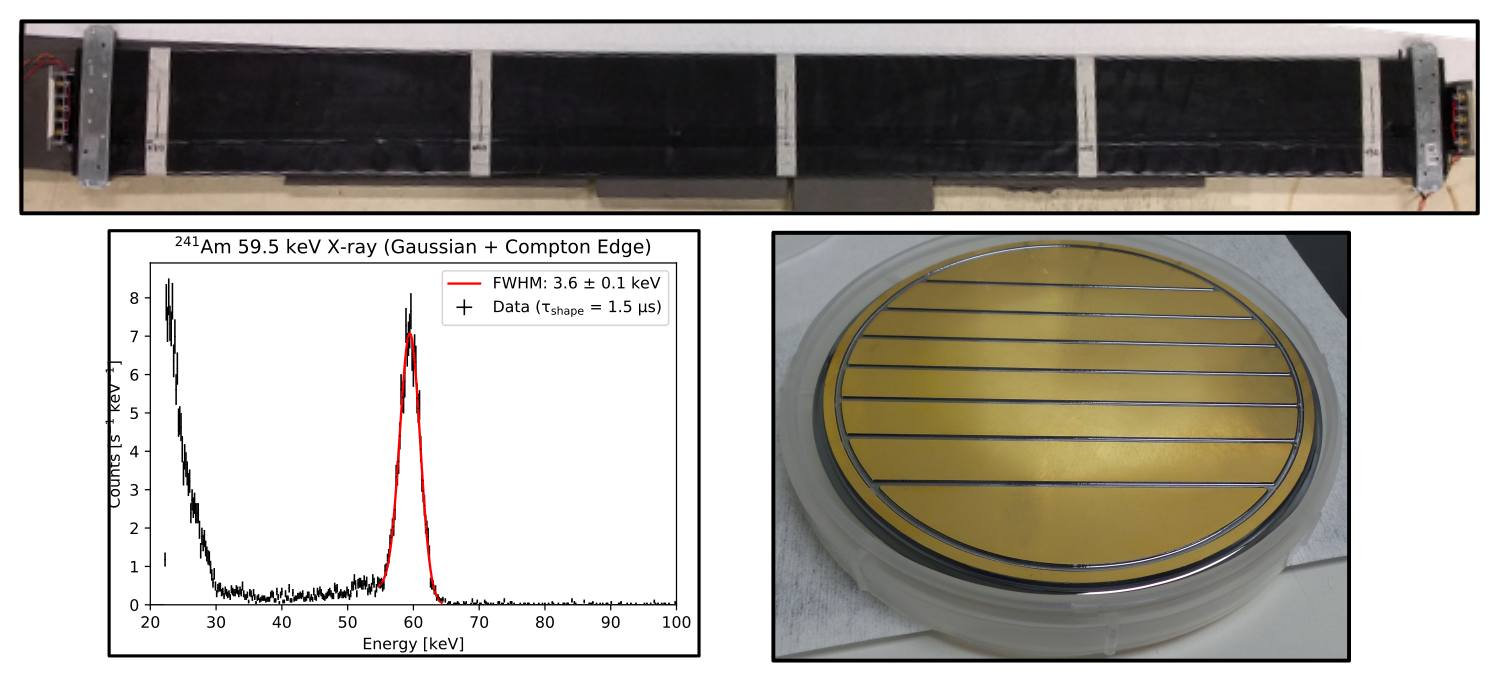}
    \caption{(Top) 1.6 m long TOF paddle with preamplifier boards integrated on the paddle ends. (Bottom left) 59.5 keV line from $^{241}$Am as measured by a Si(Li) detector with 3.6 keV FWHM energy resolution at $-40 ^{\circ}$C. (Bottom right) A bare, 8-strip Si(Li) detector.}
    \label{fig:dets}
\end{figure}

The GAPS tracker is ten layers deep, with a planned 100, 8-strip Si(Li) detectors per layer.  Each detector is 4'' in diameter and 2.5\,mm thick, providing adequate stopping power for X-rays and tracking volume for charged particles.  In order to distinguish antideuteronic X-rays from antiprotonic ones, an energy resolution of 4 keV FWHM is required.  Figure \ref{fig:dets} shows an example of a measured $^{241}$Am spectrum at $-40 ^{\circ}$C where the line width is 3.6 keV FWHM at 59.5 keV, thus satisfying the energy resolution requirement.  An oscillating heat pipe system \cite{okazaki2018} will cool the Si(Li) detectors in flight to temperatures colder than -40 $^{\circ}$C, further improving the noise performance.  A custom ASIC has been developed for reading out the Si(Li) strips that includes: a low-noise, charge sensitive preamplifier with dynamic range compression (20 keV - 100 MeV), pulse shaping and peak sampling, and analog to digital conversion.  The ASIC is in the final stages of validation, and is expected to be submitted to the foundry in December 2018.

\subsection{Time of Flight System}

The TOF detectors are 1.6 m and 1.8 m long ``paddles'' fabricated from EJ-200 plastic scintillator (Figure \ref{fig:dets}).  Each paddle end is instrumented with a preamplifier board loaded with six Hamamatsu S14160 silicon photomultipliers (SiPMs) for maximum light collection. The summed SiPM signal from each paddle end is sent to a TOF readout board, which digitizes the trace  using a DRS4 switched capacitor array ASIC.  With this configuration, the TOF system has demonstrated a $< 500$ ps end to end timing resolution in the laboratory.  In addition to measuring particle velocities and aiding in track reconstruction, the TOF system serves as the trigger to the tracker.  The current trigger implementation accepts $\sim 80\%$ of antinuclei while reducing the proton and helium rate by a factor of $10^3-10^4$.

\section{Summary and Outlook}

Antideuteron detection is a promising and unexplored probe of DM models.  GAPS is the first experiment that has been optimized to search for these rare particles, and will do so for the first time on a long duration balloon flight from McMurdo Station, Antarctica in 2020.  Follow-on flights are planned into the mid 2020's to reach a total observation time of 100 days, during which the detection of even a single antideuteron will place strong constraints on DM processes in the Milky Way.

\printbibliography
\end{document}